\documentstyle[prb,aps,multicol,epsf]{revtex}
\def \e{\varepsilon} 
\def \Q{{\bf Q}}
\def \sq{{\bf q}} 
\renewcommand{\narrowtext}{\begin{multicols}{2} \global\columnwidth20.5pc}
\renewcommand{\widetext}{\end{multicols} \global\columnwidth42.5pc}
\newcommand{\Lrule}{\vspace*{-0.2in}\noindent\vrule width3.5in height.2pt
  depth.2pt \vrule depth0em height1em}
\newcommand{\Rrule}{\vspace{-0.1in}\hfill\vrule depth1em height0pt \vrule
  width3.5in height.2pt depth.2pt\vspace*{-0.1in}}

\begin{document}

\narrowtext
{\bf Comment on PRB 59, 9195 (1999) and 62, 14061 (2000) by 
D. S. Golubev and A. D. Zaikin.} 

Golubev and Zaikin have presented calculations \cite{GZ1} which
predict a finite dephasing time at zero temperature due to the
electron-electron interaction. These calculations are in error
because ({\it i}) they do not reproduce the result of independent
calculations \cite{AAG,VA}, 
and ({\it ii}) the conclusion of ref. [\onlinecite{GZ1}] is physically
inconsistent \cite{AAG,others}.  
In their recent publications\cite{GZ2,GZS} Golubev and colaborators,
(GZS) continue to insist that the Coulomb interaction between electrons in the
regime of weak localization (WL) leads to a finite dephasing time
$\tau_{\varphi}(T)$ at zero temperature: $\tau_{\varphi} (0)=\tau_{GZS}$.  
This contribution comes from the ultraviolet part of the
electron--electron interaction and we will focus our discussion on the
ultraviolet cutoff in the theory.


\begin{figure}
\epsfxsize=0.7\hsize
\centerline{\epsffile{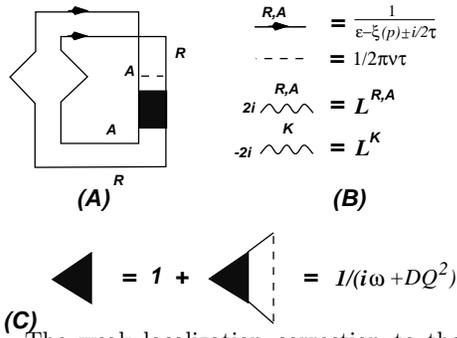}}
\caption{The weak localization correction to the conductivity is shown
in Fig.~1(A). Figure 1(B) represents the basic elements of the
diagrammatic technique. The renormalized vertex is the solution to the
diagrammatic equation, see Fig.~1(C).}
\label{Fig1}
\end{figure}

To our opinion, there is now overwhelming expermental evidence against
GZS statements.
However, we believe  that since the debate is about 
a well-defined theoretical problem, it has to be resolved by purely 
mathematical means without invoking the experimental results.

We have already published a detailed discussion of the
theoretical side of the problem\cite{AAG}. 
Taking into account the broad attention which the discussion
attracted, and the current ``antiperturbative'' sentiments among
some condensed matter theorists, we found it is worth writing 
this comment.  We highlight one of numerous mistakes in
their calculations to demonstrate that Refs.~[\onlinecite{GZ1,GZ2,GZS}] 
presented an incorrect perturbative rather than a nonperturbative treatment 
of the problem.  We address our comment to those experts in the field 
who still want to get a concrete answer to the question  
where GZS made the most crucial mistake.

\begin{figure}
\epsfxsize=0.9\hsize
\centerline{\epsffile{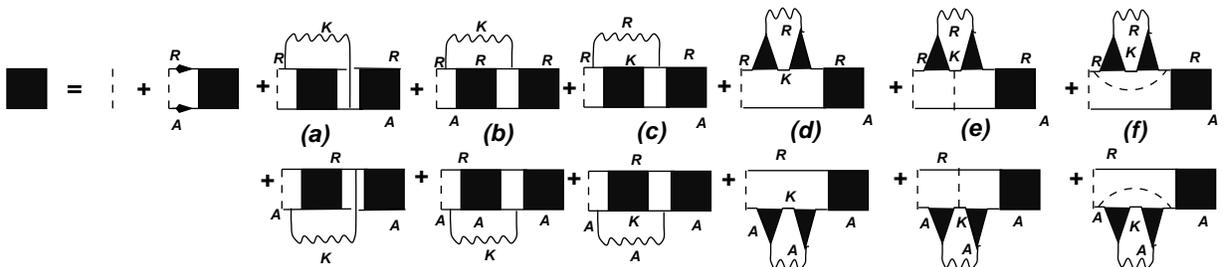}}
\caption{The diagrammatic equation for the Cooperon ${\cal C}$ in the
presence of electron-electron interaction. Note that the Cooperon and
the vertex on r.h.s. of the equation contain 
the interaction as well. The basic diagram elements
are defined in Figs.~\ref{Fig1}(B) and (C). The retarded, Keldysh and
advanced fermionic propagator are labeled by R,A, and K respectively.
In averaging procedure, one uses the relation $G^K(\e)=[G^R(\e)-G^A(\e)]
\tanh(\e/2T)$
for the exact Green functions of disordered system. }
\label{Fig2}
\end{figure}

The weak localization correction to the conductivity
\begin{equation}
\delta\sigma_{\rm wl}=-\frac{\sigma_{\rm d}}{\nu}\int
\frac{d\e d\e_1d\e_2}{(2\pi)^3}\frac{d^d\Q}{(2\pi)^d}
\frac{{\cal C}\left(
\matrix{
\e,\e_1 \cr 
\e_2,\e
}
; \Q\right)}{2T\cosh^2\e/2T}.
\label{1} 
\end{equation}
is determined by the diagram shown in Fig.~\ref{Fig1},
where the Cooperon ${\cal C}$ is defined in Fig.~\ref{Fig2}.
Here $\sigma_{\rm d}$ is the conductivity of a d-dimensional sample 
and $\nu$ is the density of states at the Fermi surface.   
Terms (a) -- (f) in Fig.~\ref{Fig2} 
describe the effect of the electron-electron interaction on the Cooperon.  
The role of terms (a) in Fig.~\ref{Fig2}  
is to cure the infrared divergence at small energy-momentum 
transfer ($Q\lesssim 1/L_\varphi$) and thus to restore the gauge
invariance \cite{AAK,AAG,VA}.
These terms do not diverge ultravioletly and thus are not important for
our discussion.  

For remaining terms (b) -- (f) we have
\begin{equation}
{\cal C}\left(
\matrix{
\e_1,\e_3 \cr 
\e_2,\e_4
}
; Q\right)=
\frac{(2\pi)^2\delta (\e_1-\e_3)\delta (\e_2-\e_4)}{-i(\e_1-\e_2)+
DQ^2+\Sigma(\e_1,\e_2,Q)}.
\label{2}
\end{equation}
\widetext
\Lrule

\Rrule
\narrowtext
The dephasing time $\tau_{\varphi}$ is determined 
by $\tau_{\varphi}^{-1}=\Sigma(\epsilon,\epsilon,Q=0 )$, 
the Cooperon's self energy at $\e_1=\e_2=\e$ and $Q=0$.  
Expansion in $\Q$ leads to the 
correction to the diffusion constant, $D$, free of ultraviolet
divergences. 
(See Ref.~[\onlinecite{AAG}] for details).

Taking only term (b) in Fig.~\ref{Fig2}  into account,
\begin{equation}
\Sigma^{\rm (b)}(\e,\e,0)={\rm Re}
\int\frac{d\omega}{\pi}\frac{d^d\sq}{(2\pi)^d}
\frac{{\rm Im} { L}^{\rm A}(\omega, \sq)}{(-i\omega+D\sq^2)}
\coth\frac{\omega}{2T},
\label{3}
\end{equation}
we encounter an ultraviolet divergence and 
$\Sigma^{{\rm (b)}}(\e,\e,0)$ 
reproduces $\tau_{\rm GZS}$ from Eq.~(71) of Ref.~[\onlinecite{GZ1}]
and Eq.~(13) of Ref.~[\onlinecite{GZS}].
We used the following relation between the Keldysh component, 
${ L}^{\rm K}(\omega,{\bf q})$, and the retarded (advanced) components 
${ L}^{\rm R(A)}(\omega,{\bf q})$ of the
screened interaction propagator: 
${ L}^{\rm K}(\omega,{\bf q})=
\coth(\omega/2T)({ L}^{\rm R}
(\omega,{\bf q}) -{ L}^{\rm A}(\omega,{\bf q})).$  
 
However, the contribution 
\begin{equation}
\Sigma^{{\rm (c-f)}}(\e,\e,0)={\rm Re}
\int\frac{d\omega}{\pi}\frac{d^d\sq}{(2\pi)^d}
\frac{{\rm Im} {L}^{\rm A}(\omega, \sq)}{(-i\omega+D\sq^2)}
\tanh\frac{\e-\omega}{2T}.
\label{4}
\end{equation}
exactly cancels out this ultraviolet divergency.
Thus, the self-energy $\Sigma=\Sigma^{\rm (b)}+\Sigma^{\rm (c-f)}$
is determined by $| \omega |\lesssim T$ 
(according to Eq.~(\ref{1}) $\epsilon \leq T$).

Comparing Eqs.~(\ref{3}) and (\ref{4}) with the result of
Ref.~[\onlinecite{GZ1}], we conclude  
the contribution of diagram (b) in Fig.~\ref{Fig2} is included in the
result of ref.~[\onlinecite{GZ1,GZS}], (see Eq.~(71) there), whereas the
contribution of diagrams (c)-(f) in Fig.~\ref{Fig2} is omitted in all
orders of perturbation theory in Ref.~[\onlinecite{GZ1}].

In the framework of GZS's approach, the contributions from diagrams (a)-(b) 
can be associated with their $S_I$, and those from (c)-(f) with their 
$i S_R$. The reason why GZ's results don't contain any contributions
from (c)-(f) is because they simply neglect the contribution of
$i S_R$ to the dephasing rate. They try to justify this (i) by
claiming in Ref.[5,6] that $S_R $ is purely real for \emph{all} paths,
and hence cannot contribute to dephasing, and (ii) by claiming in
Ref.[1,5,6] that $S_R = 0$ along classical time-reversed paths, 
and hence can be neglected
in the exponent of their path integral when the latter is
evaluated in the saddle-point approximation (in which
only classical paths are considered). 
Finally, they  claimed in Ref.[5,6] that (iii) they can 
reproduce perturbation theory by expanding
the exponent of their path integral.

All three claims are incorrect: (i) GZS obtain a real expression
for $S_R$ because in Ref. [1], because Eq.(43) of Ref.[1]
neglects Poisson brackets that are needed when
Fourier transforming the $\rho_V V^-$ and $V^- \rho_V$
terms in  Eq.(40) \cite{Jan}. If the Poisson brackets are included,
$S_R$ contains an imaginary part (which in perturbation theory
ensures that the ultraviolet divergency in $i S_R$
cancels that of $S_I$). 

(ii) GZS obtain $S_R = 0$ along classical time-reversed paths,
because in Eq.(43) of Ref.~\onlinecite{GZ1} they also [in addition to
neglecting Poisson brackets], replace\cite{footnote1} the
density matrix by its "Wigner transform",
\begin{equation}
\rho_{1'4}\stackrel{GZS}{=}\int\frac{d^3 p}{(2\pi)^3} 
e^{ip(r_{1'}-r_4)}
n\left(H\left(p,\frac{r_{1'}+r_4}{2}\right)-\mu \right),
\label{10}
\end{equation}
This replacement is unjustified, as will be shown below,
and leads them erroneously to conclude that $S_R = 0$.

Claim (iii) is false, because after neglecting
Poisson brackets, and using the ``Wigner transform Eq.~(\ref{10})'', it 
is impossible to reproduce the perturbation result in any order, 
as it will be shown below.

Let us now retrace the argument by which GZS 
claim in Ref.~\onlinecite{GZ2} that they can reproduce perturbation
theory: GZS notice correctly that the diagrams (c)-(f)
can be 
rewritten in coordinate representation as an amputated average 
[compare to Ref.~[\onlinecite{GZ2}], Eq.~(A1,A2)]
\begin{eqnarray}
2\pi\nu\tau^2
\Sigma^{\rm (c-f)}_{12}(\e,\e)&=&-{\rm Im}\int d{\bf r}_3 d{\bf r}_4
\int \frac{d\omega}{2\pi}
\label{5}
{ L}^{\rm R}_{34}(\omega)\\
&\times& \langle
G^{\rm R}_{13}(\e) G^{\rm K}_{34}(\e-\omega) 
G^{\rm R}_{42}(\e) G^{\rm A}_{12}(\e)
\rangle_{\rm amp}
,
\nonumber
\end{eqnarray}
where the Cooperon legs corresponding to the average in Eq.~(\ref{5}) 
are cut, coefficient $2\pi\nu\tau^2$ is introduced to match
with expressions (\ref{3}) and (\ref{4}), 
and $G^{\rm K}_{34}(\e)$ is the Keldysh Green function
\begin{equation}
G^{\rm K}_{34}(\e)=\tanh\frac{\e}{2T}\left[G^{\rm R}_{34}(\e)
- G^{\rm A}_{34}(\e)
\right].
\label{6}
\end{equation}
\widetext
\Lrule
\begin{figure}
\epsfxsize=0.7\hsize
\centerline{\epsffile{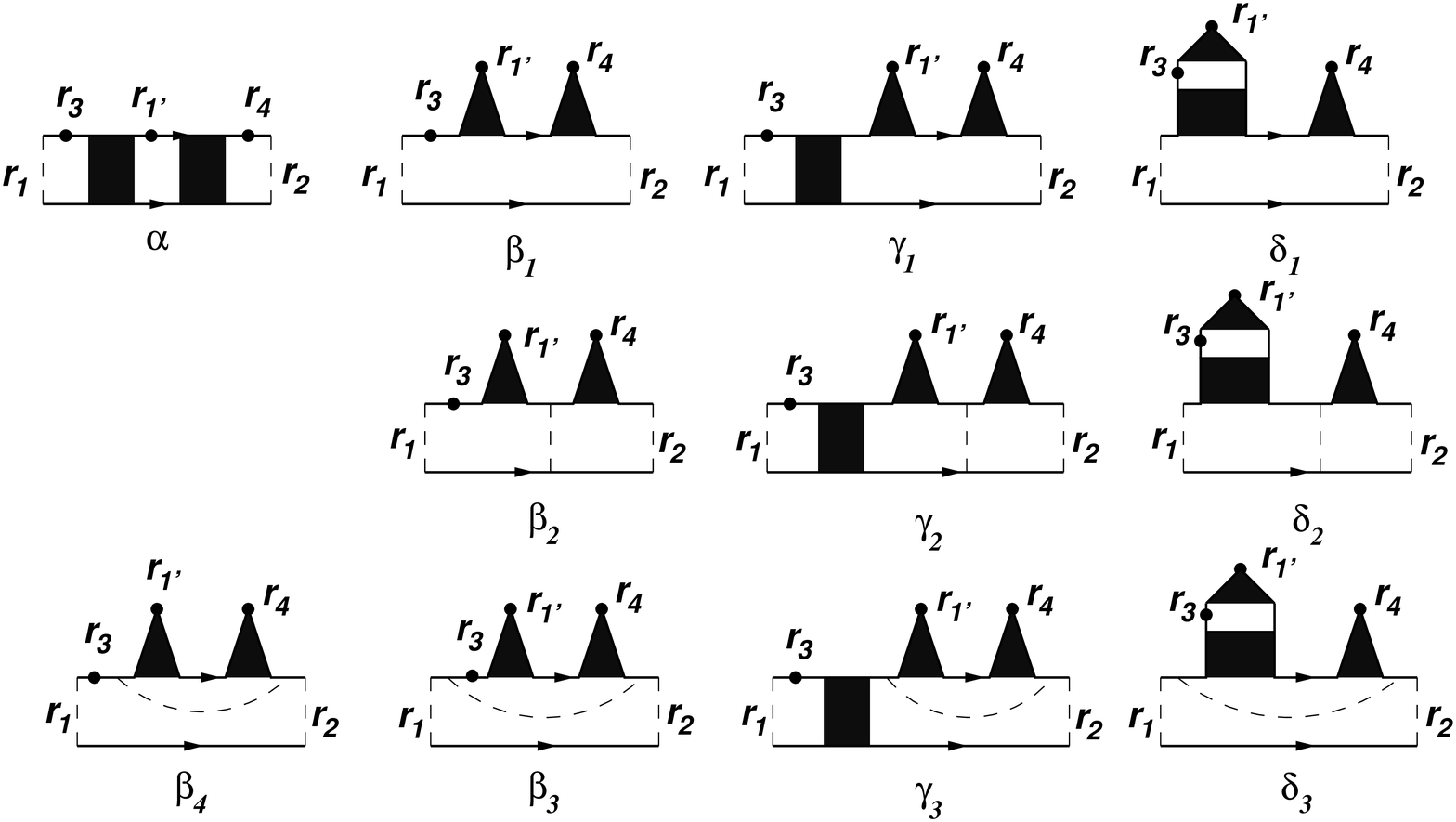}}
\caption{Diagrams for calculation of $\protect{\langle
G^{\rm R}_{13}(\e)
G^{\rm R}_{31'}(\e-\omega)[\delta_{1'4}- 2\rho_{1'4}] G^{\rm R}_{42}(\e)
G^{\rm A}_{12}(\e)
\rangle_{\rm amp}}$
}
\label{Fig3}
\end{figure}
\Rrule
\narrowtext
\noindent
GZS also  present correctly [see Ref.\onlinecite{GZ2}, Eq.~(A3)]
\begin{equation}
G^{\rm K}_{34}(\e )=\int d{\bf r}_{1'} 
\left( G^{\rm A}_{31'}(\e)-G^{\rm R}_{31'} \right)
[\delta_{1'4}-2\rho_{1'4}] 
\label{7}
\end{equation}
\noindent where $\delta_{1'4}\equiv\delta({\bf r}_{1'}-{\bf r}_4)$ and
\begin{equation}
\delta_{1'4}-2\rho_{1'4}=\int\frac{d\e}{2\pi i}G^{\rm K}_{1'4}(\e)
\label{8}
\end{equation}
is related to the density matrix $\rho_{1'4}$ 
of a non-interacting but ${\em disordered}$ system.  
Substituting Eq.~(\ref{7}) into Eq.~(\ref{5}) and using the
causality principle (analytical properties of the
integral over $\omega$) GZS obtain
\begin{eqnarray}
& &
2\pi\nu\tau^2\Sigma^{\rm (c-f)}_{12}(\e,\e)=
{\rm Im}\int d{\bf r}_3 d{\bf r}_4 
\int d{\bf r}_{1'} \frac{d\omega}{2\pi}
{ L}^{\rm R}_{34}(\omega)
\label{9}\\
& &
\times  \langle
G^{\rm R}_{13}(\e)
G^{\rm R}_{31'}(\e-\omega)
[\delta_{1'4}-2\rho_{1'4}] G^{\rm R}_{42}(\e)
G^{\rm A}_{12}(\e)
\rangle_{\rm amp}
.
\nonumber
\end{eqnarray}

Equation (\ref{9}) still contains all of the relevant 
contributions, which, if they
were calculated correctly (see below) would reproduce Eq. (\ref{4}).
It is important to emphasize, that the density matrix $\rho_{1'4}$ in 
Eq.~(\ref{9}) is a non-local operator constracted of Fermion
operators taken at coinsident moments of time but at different space
points. For a disordered system, the density matrix $\rho$ is a random
long range object; the characteristic scale after which
$\rho_{1'4}$ can be considered to be local is
$|r_{1}'-r_4| \simeq \sqrt{D/T}$. 
Since the electron Green's functions $G^{\rm R(A)}$
are also random and long range, 
the procedure of impurity averaging is not trivial,
see below.

We insist that GZS's uncontrollable approximation
Eq.(\ref{10}) of replacing the density matrix by its "Wigner
transform", which they used to argue that $S_R$ does
not contribute to the dephasing rate, is impermissible.
Once this approximation is used in Eq. (\ref{9}), it
is impossible to recover the correct result in any order of the perturbation
theory:
the "Wigner transform" neglects  all
the contributions to $\rho_{1'4}$ from all  electron trajectories
except the straight line connecting points $1'$ and $4$.  On the other
hand, those contributions are included in Eq.~(\ref{3}).  As the result of
the frivolous replacement (\ref{10}), GZS calculate two terms entering
into one physical quantity (and canceling each other),
namely $iS_R$ and $S_I$ in their language, with different 
accuracy and arrive to their  conclusions about zero
temperature dephasing!  It is neither a new physics, nor a
non-perturbative treatment - it is just an incorrect evaluation of a
perturbative contribution~(\ref{9}).

The last point we want to clarify is how to obtain Eq.~(\ref{4}) from
Eq. (\ref{9}).  A correct way is to use Eqs.~(\ref{8}) and (\ref{9})
and perform the disorder
 averaging according to standard rules, see
Fig.~\ref{Fig3}.  A straightforward calculation (see Appendix)
 gives Eq.~(\ref{4}).
Thus, the GZS's approach to introduce the density matrix into
Eq.~(\ref{5}) instead of the Keldysh Green's function
generates eleven diagrams out of original four. 
 As a result of neglecting non-locality of the density matrix and
substitution Eq.~(\ref{10}) all those diagrams are lost, i.e. scattering on
all the impurities between points 1' and 4 is not taken into account, compare
Fig.~\ref{Fig3} and Fig.~3 in Ref.~[\onlinecite{GZ2}].

\noindent

{\small
I.L. Aleiner$^{1}$, B.L. Altshuler$^{2,3}$, and  M.G. Vavilov$^4$ .
 \\
$^{1}$SUNY at Stony Brook, Stony Brook, NY 11794\\
$^{2}$Princeton University, Princeton, NJ 08544\\
$^{3}$NEC Research Institute, Princeton, NJ 08540\\
$^{4}$Theoretical Physics Institute, University of Minnesota, MN 55455
}

\section*{Appendix}
In this appendix we consider diagrams, shown in Fig. ~\ref{Fig3} and
demonstrate that Eq.~(\ref{9}) leads to Eq.~(\ref{4}). 
We insert Eq.~(\ref{8}) into Eq.~ (\ref{9}) 
to represent Eq.~(\ref{5}) in terms
of the exact Green's functions, defined for a particular disorder
realization. We have
\begin{eqnarray}
\Sigma^{\rm (c-f)}_{12}(\e,\e)
&=&
{\rm Im}\int\frac{d\omega d\e'}{i(2\pi)^2}\frac{d^d{\bf q}}{(2\pi)^d}
\tanh \frac{\e'}{2T}
{ L}^{\rm R}(\omega,{\bf q})
\nonumber\\
&\times&
\Phi_{12}({\bf q},\omega,\e,\e'), 
\end{eqnarray}
where
\begin{eqnarray}
& &
\Phi_{12}({\bf q},\omega,\e,\e')=\int d{\bf r}_3d{\bf r}_4d{\bf r}_{1'}
e^{i{\bf q}({\bf r}_3-{\bf r}_4)}\times
\\
\nonumber 
& &
\langle
G^{\rm R}_{13}(\e)
G^{\rm R}_{31'}(\e-\omega) 
[G^{\rm R}_{1'4}(\e')-G^{\rm A}_{1'4}(\e')] G^{\rm R}_{42}(\e)
G^{\rm A}_{12}(\e)
\rangle_{\rm amp}
.
\end{eqnarray}

Following the standard technique for calculation of the disorder
average $\Phi({\bf q})$ 
we obtain diagrams, shown in the Fig.~\ref{Fig3}.

The disorder averaged Green's function is 
\begin{equation}
\bar G^{\rm R, A}(\e,{\bf p}+{\bf q})=\frac{1}
{\e-\xi_{\bf p}+{\bf v}_{\rm F}{\bf q}\pm i/2\tau},
\end{equation} 
where $\tau$ is the mean free path time, 
$\xi_{\bf p}={\bf p}^2/2m-\mu$ is the energy, corresponding to
momentum ${\bf p}$, $\mu$ is the chemical potential and 
${\bf v}_{\rm F}$ is the vector of the Fermi velocity
directed along the momentum ${\bf p}$.

The diffuson ${\cal D}$ 
and the Cooperon ${\cal C}$ 
for non-interacting system (shown by black boxes) are given by 
\begin{equation}
{\cal D}(\omega, {\bf q})={\cal C}(\omega, {\bf q})
=\frac{1}{2\pi \nu \tau^2}\frac{1}{D{\bf q}^2-i\omega },
\end{equation}
and the renormalized vertex $\Gamma$ (black triangular)
is 
\begin{equation}
\Gamma(\omega,{\bf q})=\frac{1}{(D{\bf q}^2-i\omega )\tau }.
\end{equation}

Now we are ready to write down the result for diagrams in
Fig.~\ref{Fig3}. For the $\Q=0$ component of the Fourier transform 
$$
\Phi(\sq,\omega,\e,\e',\Q)=\int
\Phi_{12}(\sq,\omega,\e,\e')e^{i\Q({\bf r}_1-{\bf r}_2)}
d({\bf r}_1-{\bf r}_2)
$$ 
we obtain
\begin{mathletters}
\begin{eqnarray}
\Phi^{\alpha}_{(\Q=0)} & = &   2i\pi \nu\tau^2
\frac{1}{Dq^2-i(\e '-\e )} 
\frac{1}{Dq^2+ i\omega }
\\
\Phi^{\beta}_{\Q=0)} & = & - 2\pi \nu i \tau^2
\frac{1}{Dq^2-i(\e -\e ')} 
\frac{1}
{0 - i (\e - \omega-\e ') }
\\
\Phi^{\gamma}_{(\Q=0)} & = & 2\pi \nu i \tau^2
\frac{1}{0 -i(\e -\omega-\e ') }
\\
\nonumber
& & \times
\left[
\frac{1}{Dq^2 -i(\e-\e ')} 
+\frac{1}{ Dq^2+i\omega}
\right]
\\
\Phi^{\delta}_{(\Q=0)} & = & 2\pi \nu i \tau^2
\frac{1}{Dq^2-i(\e -\e')} 
\frac{1}{0 -i(\e - \omega-\e ')}
\end{eqnarray}
\end{mathletters}

The sum of diagrams of $\beta$ and $\delta$-type 
vanishes and for the sum of $\alpha$ and $\gamma$ type diagrams
we obtain
\begin{eqnarray}
\Phi_{(\Q=0)}
& = & \frac{4\pi \nu i \tau^2 }{0-i(\e-\omega-\e')}
{\rm Re } \left[\frac{1}{Dq^2-i(\e-\e')} 
\right]
\\
& = &
4\pi \nu i \tau^2  {\rm Re } 
\left(\frac{1}{Dq^2-i(\e-\e')} \right) \\
\nonumber
& & \times \left[
2\pi\delta(\e-\e'-\omega)+\frac{1}{-0-i(\e-\e'-\omega)}
\right].
\end{eqnarray}
The first term in the last line is just a $\delta$-function, which 
allows us to perform integration over $\e'$. The second term vanishes
after integration over $\omega$ due to analytical properties of
${ L}^{\rm R}(\omega)$ [causality principle].
As a result, we obtain Eq.~(\ref{4}).

\widetext

\begin{references}
\bibitem{Jan} We are thankful to J. von Delft for enlightening this
particular point in   GZS derivation.
\bibitem{footnote1} Compare Eqs.~(40) by Eqs.~(43)
in ref.~[\onlinecite{GZ1}], and Eqs.~(A18) and (A23) in
ref.~[\onlinecite{GZ2}]. 
See also text above Eq.~(A23) in
Ref.~[\onlinecite{GZ2}], where the difference between the exact density
matrix and its ``Wigner transformation'' is explicitely disregarded.    
See also Eq.~(29) and discussion after Eq.~(31) of Ref.~\onlinecite{GZS}, 
Hikami boxes appearing in Fig.~3 are falsely argued to vanish in
Sec.~IV of Ref.~\onlinecite{GZS}.
\end{references}
\end{document}